# Mathematical modeling of T-cells activation dynamics for CD3 molecules recycling process

Qasim Ali[1], Ibrahim H. Mustafa[2, 3], Ali Elkamel[3], Eric Touboul[4], Frederic Gruy[5], Claude Lambert[6]

[1] Department of Applied Mathematics, University of Western Ontario, London, ON, Canada.
[2] Biomedical Engineering Dept., Helwan University, Cairo, Egypt.
[3] Department of Chemical Engineering, University of Waterloo, Waterloo, ON, Canada.
[4] Henry Fayol Institute, ENSM – Saint Etienne, 158 cours fauriel, ENSM-SE – Saint Etienne, France
[5] Center for Chemical Engineering, ENSM – Saint Etienne, 158 cours fauriel, ENSM-SE, Saint Etienne, France
[6] CHU – Saint Etienne, Immunology Lab., University Hospital, F 42055, Saint-Etienne, France

## Abstract:

The down regulation of CD3 protein present on the surface of a T-cell triggers the process of activation. A rapid decrease of CD3 from the surface is required to reach the threshold that is necessary for the continuation of the activation process. By flow cytometry technique, it is well established that an amount of down regulated CD3 proteins recycle and may come on the surface of T-cell. This article is dedicated to study the impact of recycled CD3 protein that is recruited into the immune synapse at different rates. The population density of T-cells is distributed over surface protein (CD3) concentration under the framework of population balance models by using coupled system of PDEs. Such problems need non-conventional population balance modeling so that classical numerical technique can be applied. A recently developed efficient numerical method is validated against the analytical and numerical solutions.

*Corresponding author: aelkamel@uwaterloo.ca, Tel: 519-888-4567, ext. 37157, Fax: 519-746-4979

Table 1: Terminology used in this study for the definition of mathematical model.

| Variables | Description | Units |
|---|---|---|
| $B(t)$ | Activation rate of T-cells | $m^{-3} \cdot s^{-1}$ |
| $c_1(t,\tau)$ | Surface concentration of CD3 | $mol \cdot m^{-2}$ |
| $c_2(t,\tau)$ | Volume concentration of CD3i | $mol \cdot m^{-3}$ |
| $c_1'(t,\tau)$ | Surface concentration of CD3* | $mol \cdot m^{-2}$ |
| $D(c)$ | Initial distribution of non-activated T-cells | $mol^{-1} \cdot m^2$ |
| $G(c,t,\tau)$ | Reaction rate of protein concentration | $mol \cdot m^{-2} \cdot s^{-1}$ |
| $n(c,t)$ | Density of T-cells relative to surface protein | $m^{-3} \cdot mol^{-1} \cdot m^2$ |
| $N_{AC}$ | Density of activating T-cells | $m^{-3}$ |
| $N_{APC}$ | Density of Antigen Presenting Cells | $m^{-3}$ |
| $N_{NAC}$ | Density of non-activated T-cells | $m^{-3}$ |
| $N(c,t)$ | Initial distribution of activating T-cells | $mol^{-1} \cdot m^{-1} \cdot s^{-1}$ |

# 1 Introduction

T cells are responsible for the cognitive immunity leading to specific protection against infections such as virus. Each T cell clone is highly specific for one antigenic determinant that is exposed by binding of T cell receptor (TCR) with the major histocompatibility complex (MHC) molecule as a peptide – MHC complex. The TCR – peptide – MHC complex binding is very fragile but strengthened by co-receptors such as CD4 or CD8 protein (marker) and adhesion molecules such as CD3 protein (marker) forming the immune synapse (IS - a site on the surface of T-cell which is in-contact with an antigen presenting cell) [1 – 4]. As TCR does not have any cytoplasmic tail, the signal induced by the recognition is transduced through the CD3 molecule complex that is physically associated with the TCR [5 – 9]. The used CD3-TCR complex is then internalized inducing a significant decrease of the TCR-CD3 density inside the synapse present on the surface of T-cell [10,11]. A minimal level of TCR/CD3 engagement with MHC complex and internalization of CD3 in a limited period of time is required to reach an efficient threshold for a full T-cell activation [12 – 14].

The internalized CD3 (CD3i) proteins are either recycled and up-regulated to the surface of T-cell (most probably outside IS) [15,16] or are degraded and dissolved into the system due to their half-life [17 – 19]. Due to the rapid decrease of CD3 protein, T-cell recruits recycled CD3 inside the IS to continue the process of activation [16,20].

Several models have been developed to study the CD3 variability and its down regulation at the scale of single T-cell as well as population distribution of T-cells. A mathematical model of the serial triggering of TCR-CD3 complex and its down regulation was analyzed and validated against the experimental observations [10,21]. Another model investigated the serial engagement of TCR-CD3 complex and its kinetic proofreading before internalization by taking into count the half-life of TCR-peptide-MHC complex [12,22,23]. Both models studied the rapid decrease of CD3-TCR complexes and focused on the rate of internalization of CD3 without mentioning that the internalization decreases as the CD3 concentration reduces and thus recruitment is necessary [24].

CD3 recycling process has been recently modeled by presenting a non-monotonic variation in its concentration profile and distributed it over the whole population of activating T-cells by using population balances [25]. The variation is triggered by CD3 internalization and considered as the start of the activation process of a T-cell. As a continuous process it is considered that non-activated T-cells start their activation process at each time step during the whole simulation time. The model has shown couple of discrepancies which makes it less efficient. Firstly, the half-life of CD3i which is an important phenomenon during the T-cell immune response was not taken into consideration. Secondly, the limitation of the surface area covered by CD3 proteins was neglected. Therefore, the concentration of CD3 increased without indication of any bounded behavior.

This article presents a more precise model at the scale of single T-cell by considering that the surface concentration of CD3 is bounded by the maximum surface area it can attain on the surface of a T-cell. The internalized protein is down regulated either due to its half-life or it is recycled back to the surface. This shows a non-monotonic variation in its concentration. The recycling and recruitment processes of CD3 into the IS are treated separately since the recycled CD3 can be up-regulated outside synapse and requires further dynamical effects to get recruited inside IS.

The CD3 variability in T-cells is studied at the population level in the framework of population balance models (PBMs) that is represented by linear hyperbolic partial differential equation. PBMs are well established in the field of chemical engineering where the particulate processes requires an accurate predictions to understand the overall behavior of the underlying physical system [26 – 28]. In cell biology, this technique is being used frequently as cell population balance model (CPBM) where single and multi-variables cell population dynamics is studied in deterministic and stochastic settings by assimilating the chemical processes of particulate systems defined in PBM into biologically plausible expressions represented by CPBM [25,29 – 32]. The variables represent characteristics such as cell contents, size, age or biomass that is distributed over the whole population. The age-structured models are quite similar to the models normally studied in the literature. Moreover the size and the biomass also do not make a big difference since time variation in these characteristics shows a monotonic behavior as in the case of particulate processes. However, the characteristics like cell contents, e.g. protein concentration, sometimes make the problem so complex that it becomes impossible to derive the population balance equation (PBE) [25,33].

This study is devoted to describe such scenario at the population level when the time variation in the characteristic, i.e. CD3 protein, shows non-monotonic behavior and therefore affects the population distribution of T-cells. PBM is represented by coupled partial differential equations (PDEs) in order to find a population density of T-cells by using classical techniques (e.g. Method of Characteristics, Finite difference and finite volume schemes). Moreover, the population dynamics is studied by using another method which did not require a hyperbolic conservation equation, however, the method requires the reaction and activation rates (assimilated as growth and nucleation rates in chemical engineering) to find the population density function [25].

This manuscript is considerable from biological as well as mathematical modeling perspective. From biological prospective, the objective of this article is to study the impact of CD3 recruitment upon the T-cell activation process at single cell level as well as at the population level. The mathematical modeling is important as well since the behavior of characteristics is non-conventional which makes difficult to model the problem using PBMs. A new approach to deal with such problems is also validated against the classical numerical techniques.

The article is divided into four sections. In Section 2, a mathematical model is developed. In Section 3, results are presented with discussion. Section 4 concludes the whole manuscript.

# 2 Model Formulation

T-cells activation process is a complex dynamical process that triggers itself by down-regulating CD3 molecules that are present inside IS on the surface of T-cell. In this manuscript, dynamical variation in the CD3 surface concentration is modeled to study the effect of CD3 recruitment process on single T-cell as well as on population of T-cells. For single T-cell activation dynamics, CD3 degradation, recycling and recruitment processes are included in the model to observe time variation of CD3 concentration inside IS. The variation in the surface concentration of CD3 is further distributed over the whole population of activating T-cells in the framework of population balance models (PBMs). These models are defined with two biochemical processes in this study that are named as reaction and activation processes (assimilated from growth and nucleation processes in chemical engineering). The model for single T-cell activation dynamics defines the reaction rates for the reaction process while the activation rate of T-cells defines the activation process. These models are described below in detail.

## 2.1 Reaction Rates for Single T-cell

Consider the initial CD3 protein concentration $c_1 = c_{10}$ $(mol \cdot m^{-2})$ lies inside the synapse and present on the surface of T-cells. Each cell represents a population that starts its activation process at time $t = \tau$ (activation or starting time with SI unit $s$), by rapidly internalizing the CD3 protein from its surface during the whole simulation (or actual) time $t > \tau$. Assume without any loss of generality that this phenomenon follows in a continuous manner during the whole activation time $t$. The assumption is valid because T-cells have high concentration of CD3 protein in the IS which requires sufficient amount of downregulation of TCR-CD3 complex from the cell surface in a minimal duration of time during the early hours of activation, particularly first 6 hours [10]. A decreasing exponential curve can provide a good estimation of such behavior. Moreover, the internalization of CD3 continues until the T-cell get fully activated and detach itself from the APC in order to start proliferation [20]. Consequently this phenomenon provides a family of curves $c_1 = c_1(t, \tau)$ where each curve, for a fixed $\tau$, characterizes the decrease in the concentration variable of the associated population of T-cells. If $\tau$ is fixed as $\tau^*$ then the linear rate of change of CD3 can be written as,

$$\begin{cases} \dfrac{dc_1(t,\tau^*)}{dt} = -k_1 c_1(t,\tau^*), & \text{for } t > \tau^* \\ c_1(t,\tau^*) = c_{10} & \text{for } t \leq \tau^* \end{cases} \quad (1)$$

The parameter $k_1$ is the rate constant of internalization for CD3. Since the above equation is explicitly independent of $t$ and $\tau$ therefore each solution curve $c_1(t, \tau)$ starting at activation time $t = \tau$ is a translation of the first curve starting at activation time $t = 0$ such that $c_1(t,\tau) = c_1(t - \tau)$.

As the CD3 decreases on the surface due to internalization, the internalized CD3 protein (CD3i) increases its concentration $c_2$ $(mol \cdot m^{-3})$ inside T-cell with the same rate constant $k_1$. CD3i stays inside the cell for a certain time and then either dissolves by following half-life or recycles itself and come back to the surface [17,34]. In either case, the concentration of CD3i down-regulates at kinetic rate $k_2$. Thus,

$$\frac{dc_2(t,\tau^*)}{dt} = \frac{S_c}{V_c} k_1 c_1(t,\tau^*) - k_2 c_2(t,\tau^*), \quad \text{for } t > \tau^* \tag{2}$$

with initial condition $c_2(t = \tau^*, \tau^*) = 0$. Since CD3i is in the cytoplasmic fluid and CD3 is on the surface of T-cell, the conversion of the physical state was inevitable. Here $S_c$ and $V_c$ are the surface area and volume of an average T-cell. The deficiency of CD3/TCR on the surface, particularly in the synapse, compels the T-cell to recycle CD3 protein by up-regulating CD3i to the surface of T-cell by membrane transfer [15,16,35]. This recycled protein is assumed to be up-regulated outside the synapse which increases the CD3 concentration on the surface but needs to be recruited by the IS to again down-regulate during the activation process of T-cell [8]. To distinguish between the CD3 protein inside and outside IS, we use an abbreviation CD3* for the recycled protein.

If $k_3$ is the up-regulation rate constant and $k_4$ is the recruitment rate of CD3* inside the IS then the time rate of change in the recycled protein (CD3*) concentration $c_3$ ($mol \cdot m^{-2}$) on the surface of T-cell starting its activation process at a fixed time $t = \tau$ can be modeled as,

$$\frac{dc_3(t,\tau^*)}{dt} = k_3 \frac{V_c}{S_c} \frac{c_2(t,\tau^*)}{c_{10}} \left(c_{3,\max} - c_3(t,\tau^*)\right) - k_4 c_3(t,\tau^*), \quad \text{for } t > \tau^* \tag{3}$$

The parameter $c_{3,\max}$ bounds the maximum concentration of CD3* (i.e. $c_3$ cannot exceed $c_{3,\max}$). It is assumed that concentration $c_3$ is zero for $t < \tau$, i.e. $c_3(t < \tau^*, \tau^*) = 0$. The recruitment of CD3 inside immune synapse increases its concentration $c_1$. Therefore, we rewrite the Eq. (1) as,

$$\frac{dc_1(t,\tau^*)}{dt} = -k_1 c_1(t,\tau^*) + k_4 c_3(t,\tau^*), \quad \text{for } t > \tau^* \tag{4}$$

The modeled Eqs. (2) – (4) describe the recycling and recruitment process of CD3 that is presented in Figure 1.

## 2.2 Activation Rate of T-cells

Consider that $N_{APC}$ ($m^{-3}$) is the total amount of Antigen Presenting Cells (APCs) in volumetric blood stream loaded with the specific virus while $N_{NAC}$ ($m^{-3}$) is the total number of non-activated T-cells at any time $t$. A parameter $k^*$ ($m^3 \cdot s^{-1}$) is defined as the kinetic constant that indicates the volumetric flow rate of immune cells, i.e. assumed as the contact rate between non-activated T-cells and antigen presenting cells (APCs). After the recognition of virus on the surface of APC, T-cell changes its state from non-activated to the activating state, abbreviated as $N_{AC}$ ($m^{-3}$). Activating T-cells vary their protein concentration on the surface as described above which implies that the total population of T-cells would be $N_{NAC} + N_{AC}$.

It can be deduced that the rate of increase in the population of activating T-cells is equal to the rate of decrease in the population of non-activated T-cells. The rate at which T-cells change their state from "non-activated" to "activating" is termed as activation rate. For any activation time $\tau$, the activation rate can be represented by,

$$B(t) = \frac{dN_{AC}(t)}{dt} = -\frac{dN_{NAC}(t)}{dt} = k^* N_{APC} N_{NAC}(t) \tag{5}$$

The negative sign shows that the population of non-activated T-cells is decreasing. The activation rate $B(t)$ along with the reaction rates defined in previous subsection are used to study the population dynamics of T-cells.

## 2.3 Population Dynamics of T-cells

The population dynamics of T-cells is studied under the framework of population balance models (PBM) that is represented by population balance equation (PBE). The solution of PBE is a distribution of the surface concentration $c$ ($mol \cdot m^{-2}$) over the whole population of activating T-cells. In this work, it is considered that all the activating T-cells express the same initial concentration $c_0$ of CD3, therefore, the distribution of activating T-cells over the $c_0$ would be $D(c) = \delta(c - c_0)$ ($mol^{-1} \cdot m^2$). Recall that a minimal concentration of CD3 is required to be downregulated in a minimal amount of time during the early hours of activation in order to continue the process of activation. This can be said as: it is necessary to have minimal amount of CD3 concentration on the T-cell surface to continue the activation process. Thus, we consider $c_{10}$ a constant as the averaged concentration of CD3 that is necessary for the continuation of T-cell activation process. Moreover, the cell size considered in our study is constant which allows us to follow constant CD3 concentration on the cell surface. It is possible to consider random cell sizes between 4 and 11 μm so that the variation in the CD3 concentration can be proved.

The above defined phenomena provide the initial population density of activating T-cells $N(c,t)$ ($mol^{-1} \cdot s^{-1} \cdot m^{-1}$) activating at $c = c_0$ as,

$$N(c,t) = B(t)\delta(c-c_0). \tag{6}$$

At any time $t$, the initial amount of activated T-cells can be written as,

$$\int_{\Omega c} N(c,t)dc = B(t) \int_{\Omega c} \delta(c-c_0)dc, \tag{7}$$

provided that $\int_0^\infty \delta(c-c_0)dc = 1$ where $\Omega c$ is the interval representing the concentration range lies between positive real numbers. The rate of flow-in and flow-out of the population of activated T-cells between $c$ to $c+\Delta c$ is equal to the flux of the activated T-cells which travels along the characteristic curves. The slopes of these curves are denoted by reaction rates $G(c)$ ($mol \cdot m^{-2} \cdot s^{-1}$) which are defined in section 2.1. The classical representation of PBE where the source term (Eq. (6)) is treated as the boundary condition can be written as,

$$\left. \begin{aligned} & \frac{\partial n(c,t)}{\partial t} + \frac{\partial}{\partial c}\left(G(c) \cdot n(c,t)\right) = 0 \\ & n(c,0) = 0 \\ & n(c_0,t) = \frac{B(t)}{G(c_0)} \\ & \text{where } G(c) = \frac{d}{dt}c(t,\tau) \\ & \text{and } B(t) = ae^{-bt} \end{aligned} \right\} \tag{8}$$

The parameters $a = a^*b$ ($m^{-3} \cdot s^{-1}$) and $b = k^* N_{APC}$ ($s^{-1}$) are real positive numbers where $a^*$ represents the initial number of non-activated T-cells $N_{NAC}(t=0)$ interacting with APCs at the cell-to-cell interaction rate $k^*$. It is important to mention here that $G(c)$ is the implicit function of time $t$ and activation time $\tau$ while it is explicitly defined in terms of protein concentration $c$ as given in Section 2.1.

Notwithstanding PBEs are extensively used to model the particle size distributions of particulate processes and can be adapted for this study directly from the literature, it is necessary to develop the basis of PBEs for the study of biochemical processes especially when the internal variable is a protein concentration that can vary non-monotonically and distributes over the population of cells in a non-conventional manner. A direct approach to find the population density of T-cells is formulated and named as Transport Method which follows the conservation laws [25]. However, this method does not require the PBE that is represented by hyperbolic partial differential equation (PDE). The method is described briefly before the formulation of PBE for the desired problem.

## (I) Transport Method Formulation

Consider a population of T-cell activated at an activation time $t = \tau$, given in Eq. (6), varies its concentration and remains conserved throughout the course of time. In case of $t < t_{mn}$ when the curves are monotonic as shown in Figure 2, the population flows along a unique curve $c(t,\tau)$ whose slope would be

$$\frac{dc}{dt}(t,\tau) = \frac{dc}{dt}(t-\tau) \qquad (9)$$

Note that the curves are deduced by translation, i.e. $c(t, \tau) = c(t - \tau)$, while $t$ and $\tau$ are implicit time variables of the function $G(c)$ as shown in Eqs. (1), (2) and (3). Therefore, by using conservation laws, the population activated between the activation time $\tau$ and $\tau + \Delta\tau$ would be $B(\tau(c,t))\Delta\tau$ must travels along the curve $c(t, \tau)$ and remains equal to the total population between $c$ and $c + \Delta c$, i.e. $n(c,t) |\Delta c|$. This implies that $n(c,t) |\Delta c| = B(\tau(c,t))\Delta\tau$ [36]. Moreover, without any loss of generality, activation time step can be chosen as $\Delta\tau = \Delta t$. Therefore,

$$\lim_{\Delta\tau \to 0}\left|\frac{\Delta c}{\Delta\tau}\right| = \lim_{\Delta t \to 0}\left|\frac{\Delta c}{\Delta t}\right| = \left|\frac{dc}{dt}\right| \qquad (10)$$

and we can write the following equation for the population density $n(c,t)$ as:

$$n(c,t) = \frac{B(\tau(c,t))}{\left|\frac{dc}{dt}(t-\tau(c,t))\right|} \qquad (11)$$

The modulus in the denominator is taken just to avoid the negative sign since the concentration decreasing over the subsequent time follows a negative slope whereas the population density cannot be negative. Here $B$ is the activation rate constant defined in Eq. (5) and the time derivative $dc/dt$ is the reaction rate defined in previous section. Transport method (TM) provides an analytical like solution. It has been developed to study the population density function when there is only reaction and activation processes present which allows using the fact that the density of the population is varying according to the rate of change in the internal variable, i.e. concentration. This is a direct approach to find the population density function which is very useful when the internal variable, i.e. concentration, behaves non-monotonically during the course of time while the solution curves cross each other at least once [36]. The TM formulation for crossing curves is described below in more details.

Consider the two curves $c(t,\tau_A)$ and $c(t,\tau_B)$ representing the variation in the characteristic of an individual in an identically behaving population. The concentration begins to vary at activation time $t = \tau_A$ and $t = \tau_B$ (where $\tau_A < \tau_B$) that is represented by two curves crossing each other only once at time $t = t_0$ in Figure 2. In case of continuous process when there exist a curve at each activation time $\tau$, $c(t,\tau)$ is supposed to cross all its successive curves exactly once. The crossing points can be obtained either numerically using Newton-Raphson method or by interpolating $c(t,\tau)$ at each time $t$ to a polynomial and then finding the roots of that polynomial.

Particular to the model, each curve follows a translational phenomenon as the slope of each curve is same as the others. The population density can be followed from Eq. (11) for a region where there is no crossing while at a region of crossing curves, the population density can be computed by summing up their respective densities. For instance, at a crossing point $t = t_0$, $n(c,t_0)$ is the sum of $n_A(c,t_0)$ and $n_B(c,t_0)$ where $n_A(c,t_0)$ and $n_B(c,t_0)$ are the population densities of T-cells that are activated at activation time $\tau_A$ and $\tau_B$ respectively. Thus, the accumulative population density can be written as,

$$n(c,t) = \begin{cases} \dfrac{B(\tau_A(c,t))}{\left|\dfrac{dc}{dt}(t-\tau_A(c,t))\right|} & c(t,\tau_A) \neq c(t,\tau_B) \\ \dfrac{B(\tau_A(c,t))}{\left|\dfrac{dc}{dt}(t-\tau_A(c,t))\right|} + \dfrac{B(\tau_B(c,t))}{\left|\dfrac{dc}{dt}(t-\tau_B(c,t))\right|} & c(t,\tau_A) = c(t,\tau_B) \end{cases} \quad (12)$$

Each curve took the minimum value $c_{mn}$ where the slope of the curve is equal to 0 (i.e. $dc/dt|_{c=c_{mn}} = 0$) while the population density $n(c,t)$ tends to infinity (i.e. $n(c,t) \to \infty$ as $c \to c_{mn}$). However the integral of the function $n(c,t)$ remains finite (by quadrature rules), and thus the number of cells in any interval in the neighborhood of $c_{mn}$ can be computed as,

$$\int_{c_{mn}}^{c_{mn}+\Delta c} n(c,t)dc < \infty \tag{13}$$

when $c \to c_{mn}$ Numerically, in case of finite volume, the value of $n(c,t)$ at a node is a mean value that represents its integral on a grid-cell divided by its size. Thus the numerical value at the minimum concentration will be finite however it may affect the accuracy of the method.

## (II) Derivation of population balance equation

The transport method formulation can be validated in this case by deriving population balance equation when the concentration curves are following a translation phenomenon. Moreover, the interpretation of TM is quite difficult especially in multi-dynamical systems and when the initial distribution is not a dirac-delta function, contrary to the Eq. (6). Therefore, it is more convenient to adapt the classical conservation equation approach (hyperbolic equation) for the non-monotonic variation in $c(t)$. However, in some cases where the characteristic curves are not translational (i.e. each curve has its own unique slope), it becomes inevitable to model the problem in terms of classical conservation equation. For such cases, techniques like transport method helps to solve the problem [36, 37].

In case of monotonic characteristic curves $c = c(t,\tau)$ (without crossing), the conservation equation can be written as a scalar transport equation as defined in Eq. (8, with the concentration $c$ as an internal variable. However, for the crossing problem, a fixed grid is required for concentration $c$ in order to sum up the population at the crossing points. The reaction rate $G(c) = dc/dt$ is divided into two divisions $G_1(c)$ and $G_2(c)$ such that

$$G(c) = \begin{cases} G_1(c) & G(c) < 0 \\ G_2(c) & G(c) > 0 \end{cases} \tag{14}$$

The reaction rate $G_k(c)$ is then interpolated over a fixed grid of $c$. This method will avoid the crossing of curves and hence separate conservation equation would be possible for each division. The conservation equation for the population balance model for each division $k$ can be followed from system (8,

$$\frac{\partial n_k(c,t)}{\partial t} + \frac{\partial}{\partial c}\left(G(c) \cdot n_k(c,t)\right) = 0, \quad k = 1,2 \tag{15}$$

with initial conditions: $n_k(c,0) = 0$. The boundary condition for equation $k = 1$ would be $n_1(c_0,t) = B(t)/G(c_0)$. At $c = c_{mn}$ while for the second equation as $n_2(c_{mn},t) = n_1(c_{mn},t)$, where $c_{mn}$ is the minimum value of characteristic where $G(c = c_{mn}) = 0$.

The same problem arises again at the critical value ($c_{mn}$) where the slope of the concentration curve $G_k(c)$ is zero as discussed above. A way to overpass this problem is to consider a little fixed interval $[c_{mn} - \Delta c, c_{mn}]$ on which $n_1(c,t)$ is replaced by its mean value. Moreover, consider that the density $n_1(c,t)$ in the above interval is equal to the density $n_2(c,t)$ in the interval $[c_{mn}, c_{mn} + \Delta c]$ as shown in Figure 3. This is a naturally occurring case in the numerical solving using finite volume method. Nevertheless this behavior would affect the numerical accuracy.

Consider that $n \in \{n_k, k = 1,2 \mid n_{k,i}^j \approx n_k(c_i,t_j)$ is the approximate average value of $n_k(c,t)$ at time $t_j$ and a grid-cell centered at $c_i\}$. If $F^j_{k,i-1/2}$ and $F^j_{k,i+1/2}$ are the fluxes respectively upwind and downwind at the boundary of this block then the scheme takes the following form:

$$\left(n_{k,i}^{j+1} - n_{k,i}^j\right)\Delta c + \left(F^j_{k,i+1/2} - F^j_{k,i-1/2}\right)\Delta t = 0 \tag{16}$$

The fluxes are computed as upwind volumes in the following way:

$$\begin{cases} F^j_{k,i-1/2} = G(c_{i-1/2})n_{k,i-1} \\ F^j_{k,i+1/2} = G(c_{i+1/2})n_{k,i} \end{cases} \tag{17}$$

Each equation is solved on the same grid of concentration between $c_{mn}$ and $c_0$ so that the population densities at the crossing point could be added. For this reason the growth function is interpolated over a fixed concentration grid. The principle of the scheme is exposed in Figure 3. This scheme, known to be stable provided that the classical CFL condition is satisfied on the time step, presents the drawback of developing a high numerical diffusion. In order to decrease the numerical diffusion around the discontinuity front, the flux on the boundary between two grid blocks can be set to 0 until this boundary is reached by the front computed from $G(c)$. But this method does not prevent diffusion behind the front. The result derived is then very close to the solution found by Eq. (12).

## 2.4 Parameterization

The kinetic rates defined for CD3 dynamical variability on the cell surface and its dissolution inside the T-cell are obtained from the empirical results present in the literature, particularly in case of influenza virus infection, [8,10,20,24,37 – 40]. All the parameters are defined in Table 2. The CD3 molecules concentration varies according to the size of the cell. In this article, the values of the surface ($S_C = 3 \cdot 10^{-10}$ $m^2$) and the volume ($V_c = 5 \cdot 10^{-16}$ $m^3$) of T-cells are defined for 5 μm radius while the contact area between the surface of a T-cell and APC (immunological synapse - IS) is considered as the 25% of the T-cell surface area, i.e. $7.5 \cdot 10^{-11}$ $m^2$ [38,41]. The CD3-TCR binding also varies and depends upon the affinity of T-cell receptor. The initial concentration of CD3-TCR molecules are considered to be $c_{10} = 3.24 \times 10^{-10}$ $mol/m^2$ which represents around 30,000 molecules inside an immune synapse (IS) to be ready to bind and get internalized into T-cell cytosol (intracellular fluid). It is assumed that rest of the surface of T-cells can achieve the maximum concentration of CD3-TCR complex as same as the concentration present inside IS. Thus, $c_{3,max} = c_{10}$.

The parameters defined for the activation rate $B(t)$ depends upon the total number of APCs and T-cells present per unit of time per unit of volume and the volumetric contact rate $k^*$ between them. Since each APC can interact with several T-cells at a particular time, therefore it is more convenient to count the APCs contact sites, i.e. $N_{APC} = 4 \times 10^{12}$ cells/m$^3$. The initial density of non-activated T-cells is chosen quite close to total sites available on APCs ($N_{NAC}(0) = 2.16 \times 10^{12}$) while $N_{NAC}$ decreases with time as soon as they interact with APCs. The mobility rate of cells is chosen less than 1 μm / sec, i.e. $k^* = 1/3 \times 10^{-16}$ [42]. Thus, the total interactions between $N_{NAC}$ and $N_{APC}$ can be defined as $a = k^* \cdot N_{NAC}(0) \cdot N_{APC} = 3 \times 10^8$ $m^{-3}s^{-1}$ whereas the contact rate of $N_{APC}$ can be written as $b = k^* \cdot N_{APC} = 1/7200$ $s^{-1}$

The kinetic parameter responsible for the internalization of CD3 proteins depends upon various parameters including the contact area between cells, association/dissociation rates of receptor markers, TCR-peptide/MHC affinity and their concentrations [43 – 45]. Experimental evidence from the literature shows that the CD3 concentration decreases rapidly in the first few hours which help T-cells to complete their activation process [20]. The rate of internalization is chosen as $k_1 = $ $1/3 \times 10^{-3}$ $s^{-1}$ which sufficiently decreases the CD3 concentration in first 5 hours [10]. The diminution rate of internalized CD3 (CD3i) is an important parameter in this model since degraded CD3i recycles or dissolves into the system. In addition, the CD3i motivates T-cell to continue its process of activation and produce further protein, e.g. CD69 and CD25 [33]. The baseline value for the half-life/diminution of internalized CD3 is chosen $k_2 = 1/3 \times 10^{-4}$ $s^{-1}$ [37,39,40,46]. The kinetic rate for recycling of CD3i is chosen $k_3 = 6 \cdot 10^{-5}$ $s^{-1}$ while the recruitment rate of CD3* protein is varied to see how it affects the overall concentration of CD3 inside IS. A realistic parametric space is defined to investigate the variability of CD3 over the surface of T-cell [20,45] whereas a parametric analysis is considered mandatory to study parametric effects upon the CD3 recycling process.

Table 2: Parametric values that are used in the results

| Abb. | Description (without units) | Values | Units |
|---|---|---|---|
| a | Number of Contacts between $N_{APC}$ and $N_{NAC}$ | $3 \times 10^8$ | $m^{-3} \cdot s^{-1}$ |
| a* | Initial density of non-activated T-cells | $2.16 \times 10^{12}$ | $m^{-3}$ |
| b | contact rate of $N_{APC}$ | 1/7200 | $s^{-1}$ |
| k* | Volumetric flow (cells mobility) rate | $1/3 \times 10^{-16}$ | $m^3 \cdot s^{-1}$ |
| $k_1$ | rate of CD3 internalization | $1/3 \times 10^{-3}$ | $s^{-1}$ |
| $k_2$ | Rate of CD3i diminution | $1/3 \times 10^{-4}$ | $s^{-1}$ |
| $k_3$ | Rate of recycling CD3i | $6 \times 10^{-5}$ | $s^{-1}$ |
| $k_4$ | Rate of recruitment of CD3* | $6 \times 10^{-5} - 6 \times 10^{-8}$ | $s^{-1}$ |
| $N_{APC}$ | Density of Antigen Presenting Cells | $4 \times 10^{12}$ | $m^{-3}$ |
| $t$ and $\tau$ | actual time and activation time | $10^6$ | $s$ |
| $S_c$ | Surface area of T-cell | $3 \times 10^{-10}$ | $m^2$ |
| $V_c$ | Volume of T-cell | $5 \times 10^{-16}$ | $m^3$ |

# 3 Results and Discussion

The above described models are solved by using numerical schemes and validated against analytical methods where possible. In case of no possibility of analytical solution, the numerical results are compared with transport method to observe their commitment with each other. At scale of single T-cell, the system of Eqs. (2), (3) and (4) are solved simultaneously. At the population scale, the analytical solution was found by using Method of Characteristics (MOC) where the slope of the characteristic is followed from the equations (2) – (4). The numerical methods used in this section are finite volume (FVM) and the numerical Transport Method (TM) [25,37].

## 3.1 Single T-cell dynamics

The solutions found for the system of Eqs. (2) – (4) provide concentrations of internalized CD3 ($c_2$), recycled CD3 ($c_3$) and CD3 inside IS ($c_4$). At fixed activation time $\tau$ and by assuming recruitment rate $k_4 = 0$, it is possible to find the analytical solution of the system. However the use of CD3 recycling process is not worthy to mention in the model without the recruitment of CD3*, therefore it is important to consider $k_4$. This necessity can be clearly observed in Figure 4 where CD3 variation is shown inside the IS in (a) as well as over the whole surface of T-cell (CD3 + CD3*) in (b). When $k_4 = 0$, the CD3 concentration directly approaches zero while a gradual increase in $k_4$ stabilizes the CD3 inside IS (see Figure 4(a)). One can also observe that the high values of recruitment rate (e.g. $k_4 > 6 \cdot 10^{-8}$) defines a lower limit so that the concentration of CD3 inside the IS remain constant after time $t$. This is possibly due to the balance between the recruitment and the internalization terms of CD3 in Eq. (4). On the other hand, in Figure 4(b), the CD3 concentration on the overall surface of T-cell ($c_1 + c_3$) decreases to the same level for all values of $k_4$ and provides a lower bound. Afterwards, the concentration continuously increases and it shows very similar behavior for the parameter values $k_4 < 10^{-6}$. When the recruitment rate is increased ($k_4 > 10^{-6}$), the overall surface concentration of CD3 still increases however the slope of the solution curve $c_1 + c_3$ decreases rapidly. In case of $k_4 = 6 \cdot 10^{-5}$, the slope approaches zero after some $t \approx 3 \cdot 10^4$. This indicates a balance among internalization, recycling and recruitment processes of CD3 so that CD3 remain constant on the surface of T-cell.

For the recruitment rates, $k_4 = 6 \cdot 10^{-5}$ and $k_4 = 6 \cdot 10^{-8}$, the kinetic equations (2) – (4) are solved numerically by using Runge-Kutta method of order 4 (RK-4) as shown in Figure 5 and Figure 6. Since the equations are explicit functions of only protein concentrations (not of time $t$ or $\tau$), therefore the slope of each curve is same for all the activation time $\tau$. The multiple curves in each subplot are drawn by replacing $t$ by $t - \tau$ in the system (2) – (4). It is important to mention here that the activation time step $\Delta\tau \to 0$ in the model simulations whereas each subplot in the figures show only 10 curves so that one can observe their translatory behavior. Moreover, the crossing of the curves, particularly in Figure 5d and Figure 6d, form continuous portions in the subplots in which any two T-cells who start their activation process at different times cross each other while one is increasing and the other is decreasing its respected protein concentration.

The CD3 concentration ($c_1$) shows an exponential decrease from the IS in the first few hours of T-cell activation which gradually slows down due to the deficiency of CD3 as shown in Figure 5a (less rapidly) and Figure 6a (more rapidly). Similar results have been found in the literature by monitoring experimentally initial five hours of down-regulation of TCR-CD3-ζ complexes [10]. The down-regulation of CD3 continues until the T-cell becomes fully activated. In response, the internalized CD3, i.e. CD3i, protein rapidly increases inside the cell at the beginning and then starts decreasing due to low concentration of CD3 internalization and its diminution inside the cytosol (by recycling or dissolution) [17], see Figure 5b and Figure 6b. A continuous internalization of CD3 is necessary to keep compelling the T-cell to continue its activation process by producing several other intracellular and extracellular signals (proteins). At this point recycling of CD3i protein helps to maintain the sufficient amount of CD3 on the cell surface that is followed by its recruitment to the IS.

The T-cell induces CD3 protein either by recycling or by producing more CD3 molecules and presents them on its surface. In this study, it is assumed that this recycled protein, CD3*, lies outside the IS. CD3* concentration increases rapidly in the beginning and continuously follows an increase during the whole simulation time. However, the slope of the curve decreases with time because the recycled protein is recruited at the same time to the IS with recruitment rate $k_4$, see Figure 5c and Figure 6c. The time variation in the total concentration of CD3 on the surface of T-cell (CD3+CD3*) shows that the decrease in the CD3 molecules during the first few hours is overcome by its recycling and recruitment in the rest of T-cell activation process as shown in Figure 5d and Figure 6d. The balance in the recruitment and degradation of CD3 can be well observed after $t = 25000$ $s$ (approx. 7 hours) in the Figure 6b and Figure 6d.

The solution of the system (2) – (4) are inserted into the system (8) and treated as the reaction rates of the T-cells in the cell population balance equations, i.e. linear hyperbolic partial differential equations (PDEs) as discussed below.

## 3.2 Population dynamics of T-cells

T-cells population dynamics is studied for the surface protein concentrations since the flow-cytometry analysis counts the T-cells only by means of surface markers [20,47]. Therefore population density of activating T-cells is studied for CD3 protein concentration, $c_1$, which lies in the IS and for the sum of CD3 and CD3* proteins concentration, $c_4$, which represents the total CD3 concentration on the surface of T-cell. The analytical solution is found by MOC for $c_1$ and numerical solution by Upwind Scheme and Transport Method (TM) for both $c_1$ and $c_4$. The results are presented in Figure 7, Figure 8 and Figure 9 whereas multiple curves in each figure represent actual times that are mentioned respectively.

The analytical solution of CD3 T-cells population dynamics is only possible when there is no recruitment term, i.e. $k_4 = 0$ (see Figure 7). All the T-cells begin their activation process with the same CD3 protein concentration ($c_1$) on their surfaces, however the density of T-cells starting their activation process decreases with the increase in time. This can be observed from the Figure 7 that population of T-cells starts their activation process at a certain activation time $\tau$ by decreasing CD3 protein from the immune synapse (IS). At the next time step, few of the other T-cells come in-contact with the APCs and start following the process of activation. However, the boundary condition suggest that the preceding density of T-cells is higher than the next population which can be observed from the right end of the curve in Figure 7. This process continues during the whole simulation time so that a population of T-cells left with a very small concentration of CD3 while very small population of T-cells has high concentration.

The importance of CD3 recruitment can be explored by investigating downregulation rate of CD3 protein when CD3-TCR complexes are serially engaged with MHC-peptides. In case of rapid CD3 internalization, T-cells may not have enough CD3 concentration present in the IS after first few hours to continue the activation process. This phenomenon is quite natural since the activating T-cells, in Figure 7, have no recruitment term and eventually huge number of T-cells may not be able to follow the activation process due to lack of CD3 concentration. However, the previous studies reveal that most of the T-cells who successfully complete their first few hours of activation make it possible to complete their activation process. This means that a minimal level of CD3 concentration remains in the IS after the crucial period of time which allows T-cells to remain in contact with APCs until they get fully activated. Thus a certain amount of recruitment is necessary to maintain the minimum level of CD3 concentration for the full activation of T-cells.

The recruitment of CD3 in the activating T-cells shows an appealing behavior in the population dynamics when a discontinuity occurs in the middle of the distribution function. In our model, this discontinuity occurs after the first population of activating T-cells achieve the minimum level of CD3 concentration on their surfaces. Moreover, the discontinuity divides the population density of activating T-cells in such a way that one portion of distribution curve raises due to the overlapping populations while the other portion remain same as shown in the Figure 8 and Figure 9. The portion of overlapping populations is a sum of two populations: first is the one whose CD3 internalization is higher than the combined effect of recycling and recruitment and therefore decreases the concentration of CD3 from the T-cells surfaces while the second population is the one whose CD3 concentration is increasing after attaining the minimum level of CD3 inside IS.

The more interesting phenomenon is the movement of discontinuity from left to right of the curve with the increase in actual time '$t$' which shows that the populations are overlapping each other at high concentration levels of CD3. At low rate of recruitment, the recycled protein gathers outside the IS and therefore the total concentration of CD3 on the surface of T-cells increases however the internalization remains slow. On the other hand, the high rate of recruitment allows more recycled protein to join IS and therefore the internalization of CD3 remains high. Thus, the discontinuity moves to further right when the recruitment rate is low as shown in Figure 8 and Figure 9. Moreover, the curve that is at the left of the discontinuity remains smooth since the crossing is observed continuously in this portion.

The recruitment of CD3 in the immune synapse is inevitable however it is important to observe the population behavior at different kinetic rates of CD3 recruitment which can provide more insight to its impact upon the T-cell activation process. It can be observed from the figures that the T-cells population is dense in the range of high CD3 concentration when the recruitment rate is slow (see Figure 9). Moreover, the population density is less diverse in case of slow recruitment rate as one can look at the left discontinuity in the Figure 8 and Figure 9. The reason could be simple as that the fast recruitment of CD3 to IS permits the T-cells to engage more CD3 with APCs and degrade more of them which consequently keeps high density of T-cells at the low concentration region of CD3 (see Figure 8). Moreover, recruitment of CD3 can be affected by the recycling of CD3i due to its dependent behavior.

# 4 Conclusion

CD3 recycling process is an important physiological phenomenon since it is necessary throughout the T-cells activation process against viral/bacterial infections which involves antigen presenting cells. The multi-scale population balance modeling has portrait a clear image of the impact of CD3 recycling and recruitment processes upon the population density of T-cells. Apparently huge amount of cells seem to have medium concentration of CD3 protein (see Figure 8 and Figure 9), however there are actually two populations which are at totally different states. One begins the activation process and leads to a decrease in the CD3 protein concentration while the other one is at the end of its activation process and ready to divide in next few hours as it can be observed in Figure 5 and Figure 6.

The article focused on the modelling portion because, in the literature, the (cell) population balance modelling is studied only for monotonically increasing/decreasing characteristic of the population density function however this article focuses on a non-monotonic increase when the characteristics are crossing each other. In chemical engineering, the characteristics are chosen as particle age, length, volume and porosity. In biochemical engineering, the characteristics are cell size, age and biomass. These characteristics are always chosen as increasing/decreasing (monotonic) functions of time during the whole simulation. Thus, it was not worthy for the previous modellers to describe the model in detail with figures.

This article focused on the non-monotonic variations in the characteristic (i.e. CD3+CD3* concentration on the cell surface) and therefore provide details to make the basis for population balance equation (PBE) in case of crossing characteristics. It has provided an extension of the article where PBE was not possible to be modelled by using the classical representation of the conservation equation (please see [36]). The mathematical modelling of PBE became possible under non-monotonic behavior of the characteristic when the succeeding curve at each time step was considered as the translation of its any previous curve. Moreover, this article provides the basis to the modelling of cell population balance equation which depends upon surface protein concentration, where only reaction and activation terms are present (growth and nucleation in terms of chemical engineering).

Transport method was presented for finding the population density function without using classical conservation equation. In case of crossing, when the analytical solution is not possible, TM provides analytical like solution. At the same time it is quite easy to follow this method and it takes very less simulation time as compared to classical numerical methods. There is a possibility of extension of this method by making continuous distribution of $D(c)$ and including several other intracellular/extracellular reactions where numerical solutions could be more difficult to follow, e.g. when an extracellular protein like IL-2 binds to its receptor CD25 [36, 37].

# Disclosure

The authors declare that there is no conflict of interests regarding the publication of this paper.

[45] S. Valitutti, D. Coombs, L. Dupré, FEBS Letters 2010, 584, 4851.

[46] S. N. Arkhipov, I. V. Maly, PLoS ONE 2007, 2, e633.

[47] A. L. Givan, Flow cytometry: first principles, John Wiley & Sons 2013.24

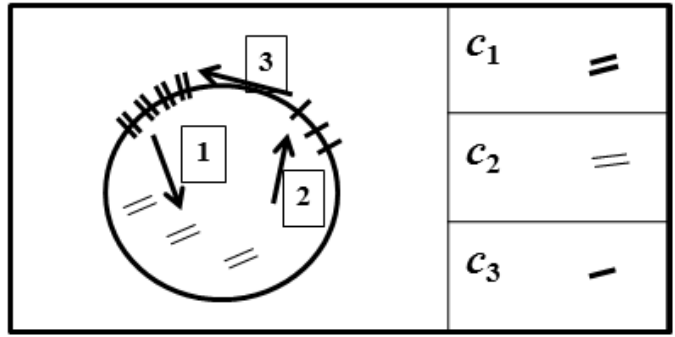

Figure 1: Three steps of CD3 molecules dynamical process. 1: Decrease in $c_1$ (CD3 concentration inside immune synapse (IS)) inducing the increase in $c_2$ (CD3i concentration) due to CD3 internalization. 2: Recycling of CD3 induces $c_3$ (CD3*). 3: Recruitment of $c_3$ into the IS.

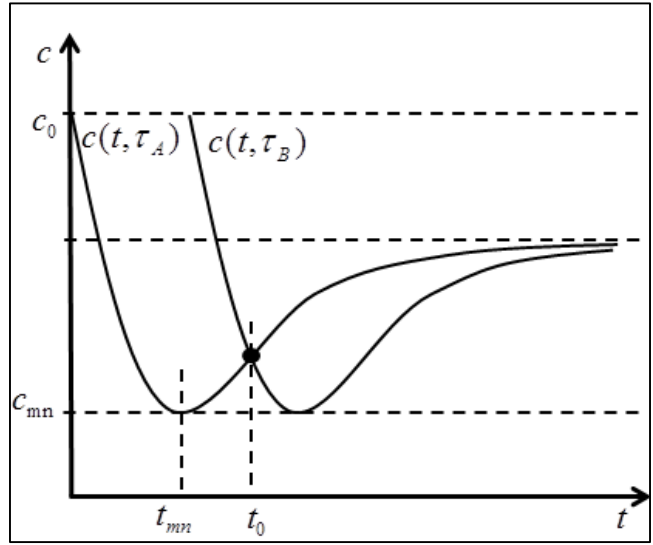

Figure 2: Two characteristic curves start at initial concentration $c_0$ at activation times $\tau_A$ and $\tau_B$ and attains a minimum value $c_{mn}$ at their respective times and crosses each other only once at time $t_0$.

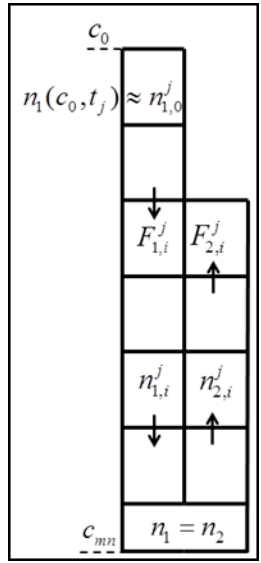

Figure 3: Principle of the numerical upwind scheme used in this study to find the population density function. In this figure, $c_0$ is the initial concentration while $c_{mn}$ is the minimum concentration where it is assumed that $n_1 = n_2$.

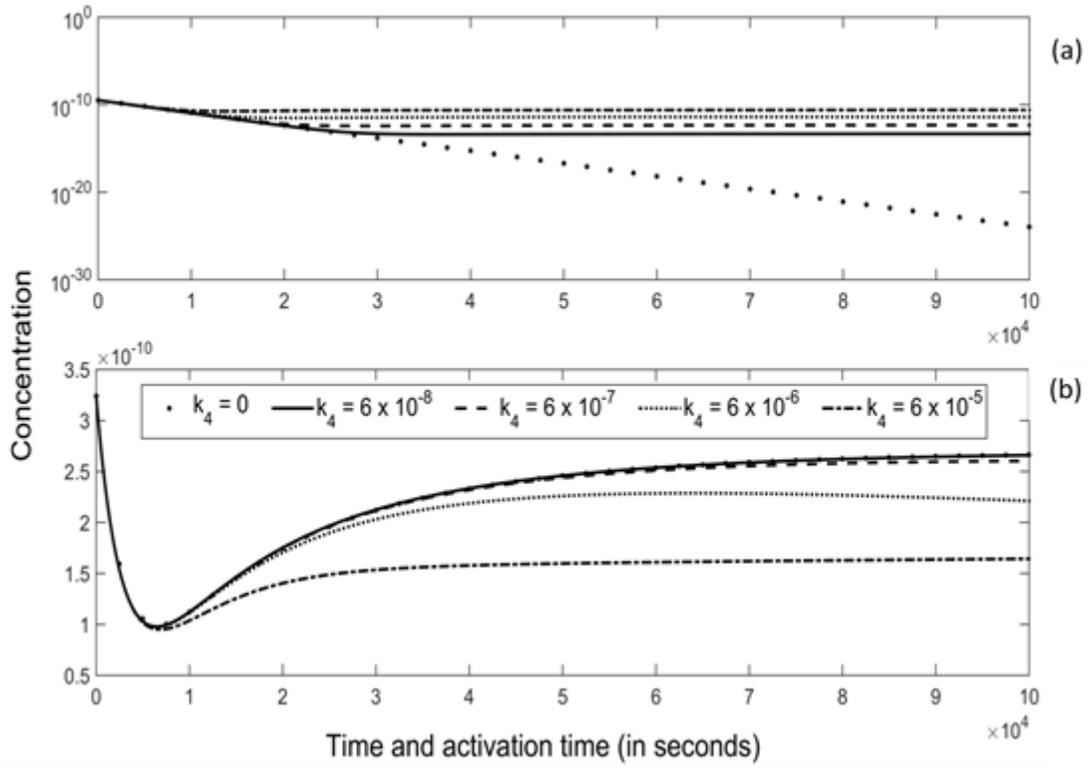

Figure 4: Effect of recruitment upon CD3 protein concentration inside IS (a) and over whole T-cell surface (b) at five subsequent values of the recruitment rate $k_4$. All the other parameters are defined in Table 2.

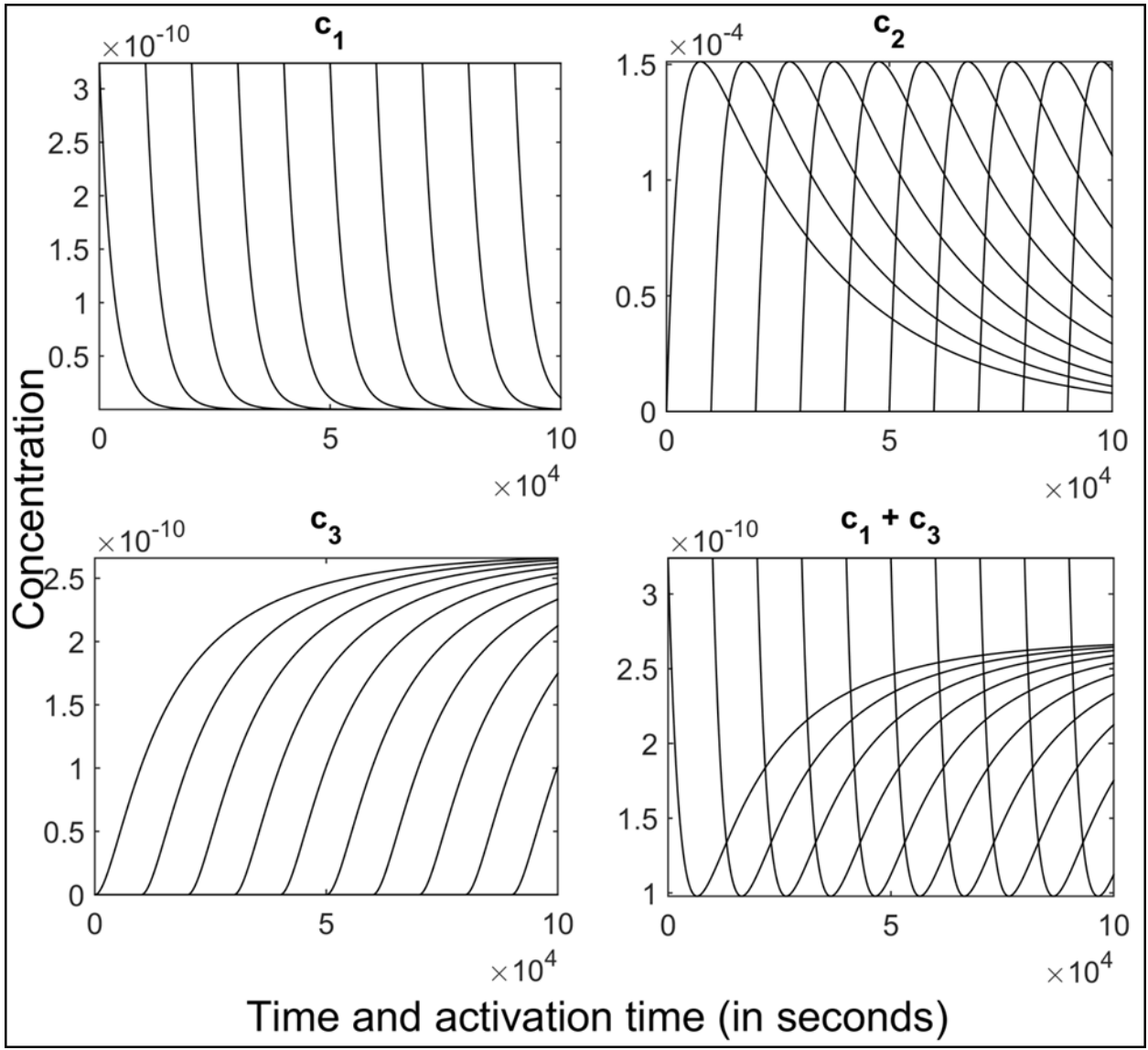

Figure 5: Numerical solution for single T-cell dynamics of CD3 protein concentration with slow recruitment of the recycled protein in the immune synapse. The kinetic parameters are $k_1 = 1/3000$, $k_2 = 1/30000$, $k_3 = 6 \times 10^{-5}$, $k_4 = 6 \times 10^{-8}$. Horizontal axis is the time $t$ and activation time $\tau$ in seconds while vertical axis is the concentration $c$ (mol/m$^2$ for plot (a), (c) and (d), and mol/m$^3$ for plot (b))

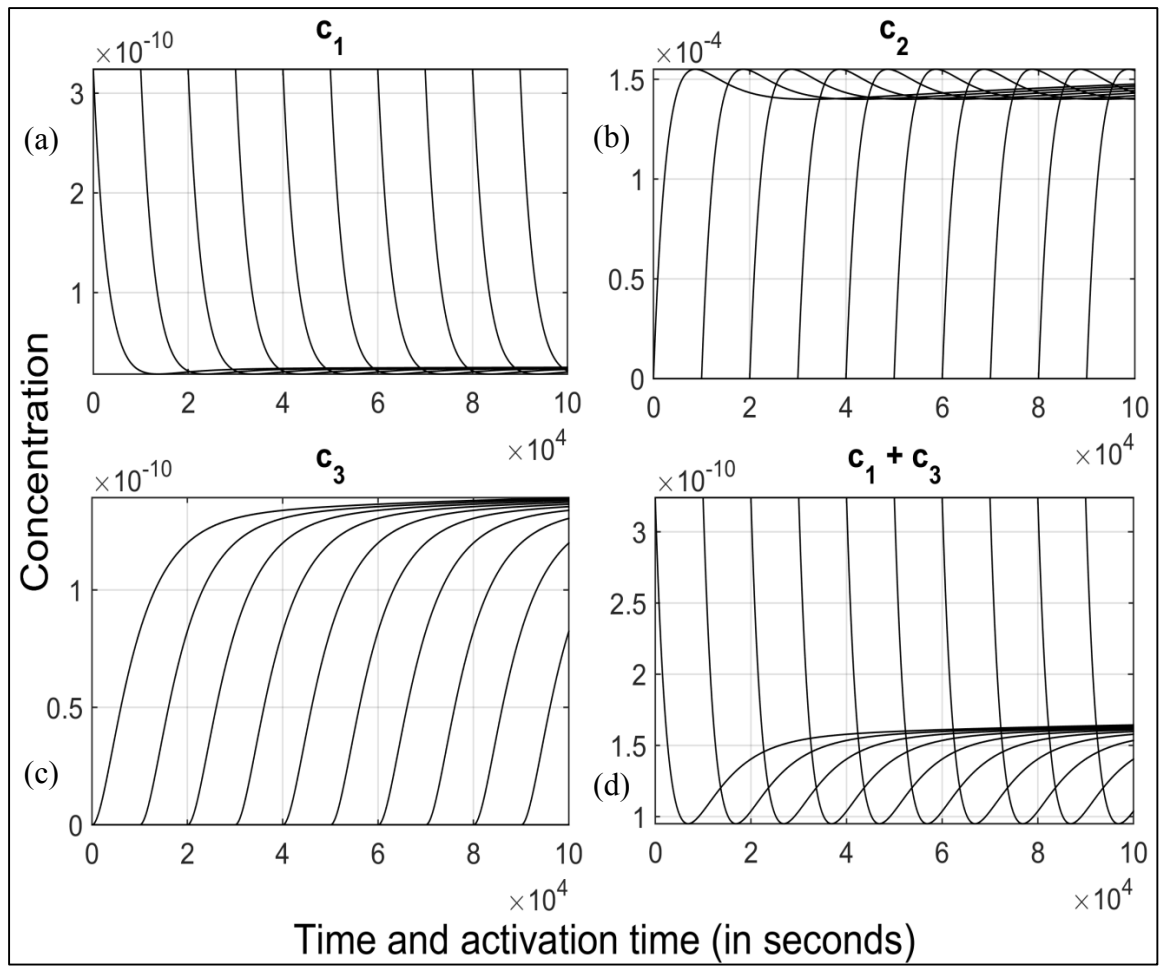

Figure 6: Numerical solution for single T-cell dynamics of CD3 protein concentration with fast recruitment of the recycled protein in the immune synapse. The kinetic parameters are $k_1 = 1/3000$, $k_2 = 1/30000$, $k_3 = 6 \times 10^{-5}$, $k_4 = 6 \times 10^{-5}$. Horizontal axis is the time $t$ and activation time $\tau$ in seconds while vertical axis is the concentration $c$ (mol/m$^2$ for plot (a), (c) and (d), and mol/m$^3$ for plot (b)).

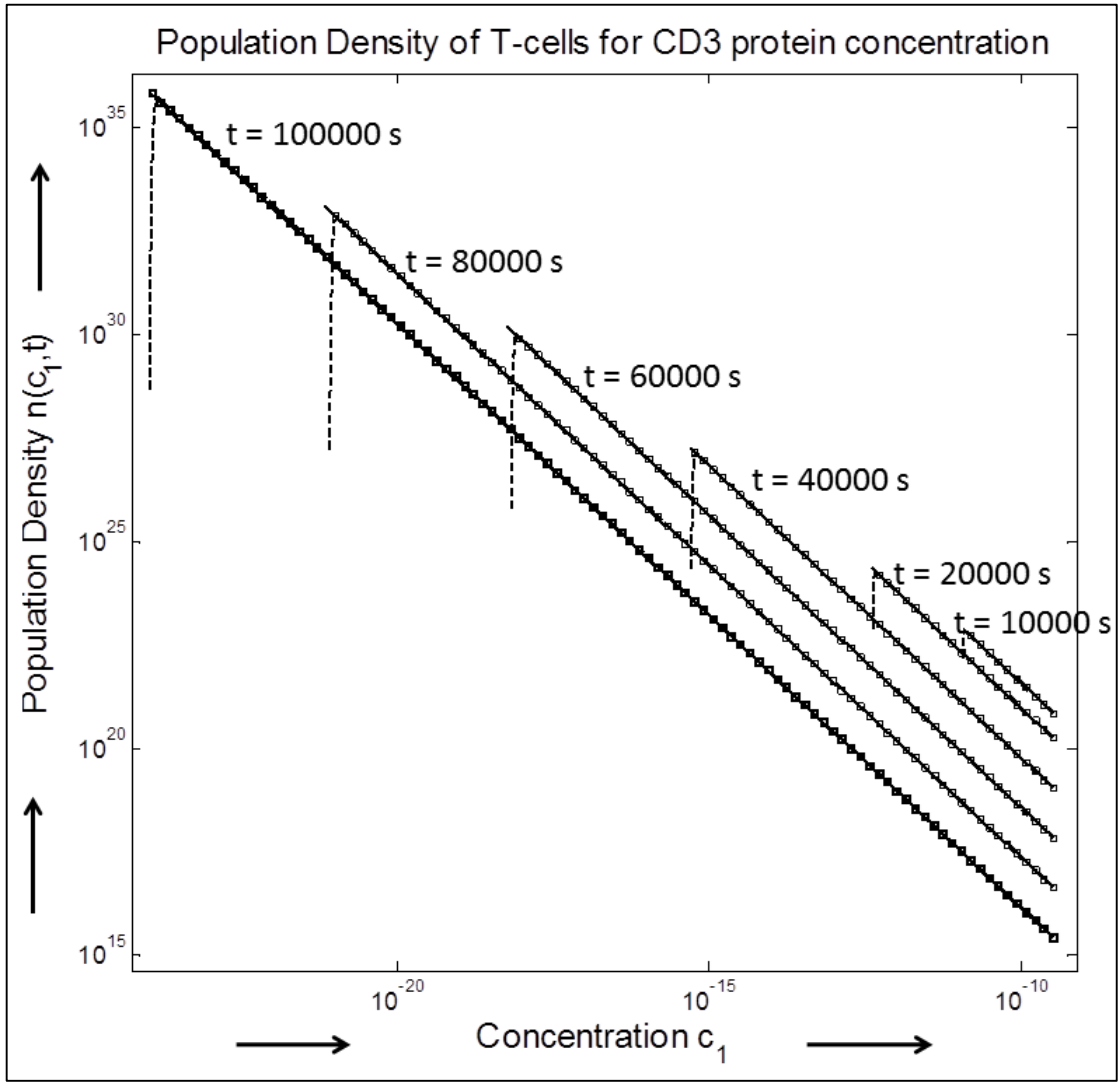

Figure 7: Population density of T-cells against CD3 concentration at six different times without any recruitment of CD3. The solution is found by three methods: MOC (squares □), TM (Lines −) and FVM (Dashes --). All the parameters are defined in Table 2 except $k_4 = 0$.

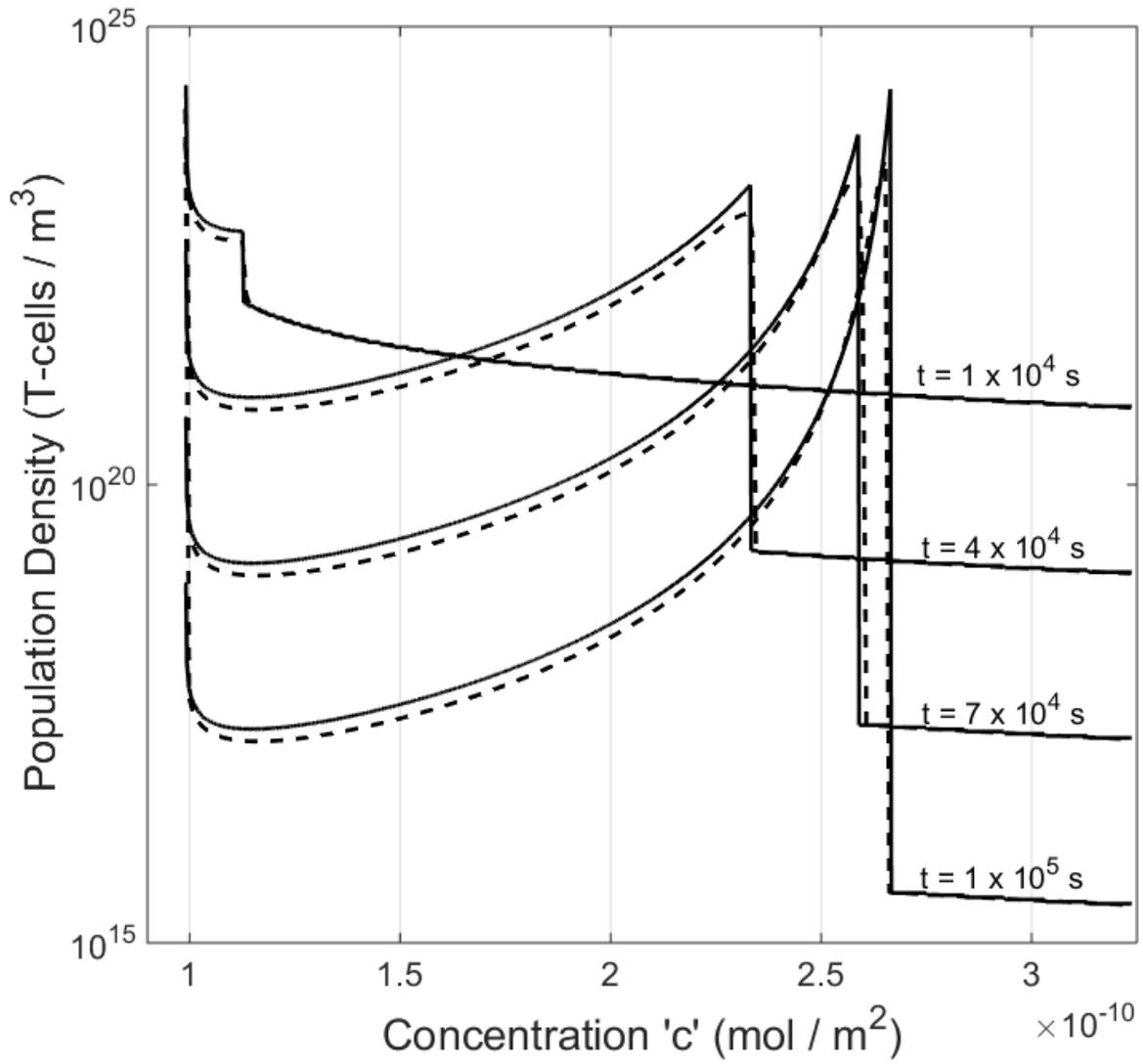

Figure 8: Population density of T-cells for CD3 +CD3* T-cells at four different actual times with slow recruitment rate $k_4 = 6 \cdot 10^{-8}$. All the other parameters are defined in Table 2. The solution is found by two numerical methods: TM (Lines −) and FVM (Dashes - -). The vertical axis is the population density $n(c,t)$ while the horizontal axis is the concentration of CD3 + CD3*, i.e. $c = c_1 + c_3$.

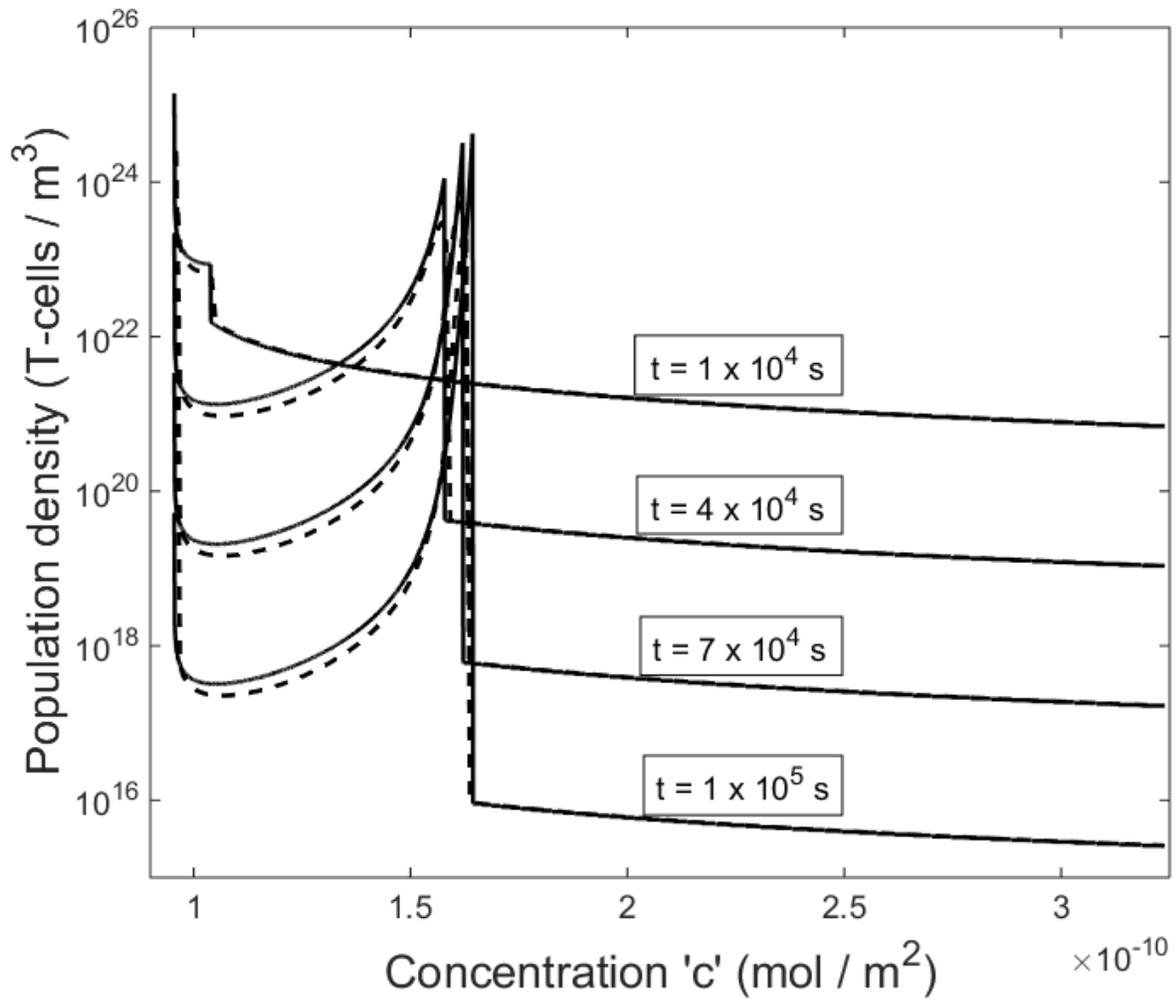

Figure 9: Population density of T-cells for CD3 +CD3* T-cells at four different actual times with fast recruitment rate $k_4 = 6 \cdot 10^{-5}$. All the other parameters are defined in Table 2. The solution is found by two numerical methods: TM (Lines − ) and FVM (Dashes - -). The vertical axis is the population density $n(c,t)$ while the horizontal axis is the concentration of CD3 + CD3*, i.e. $c = c_1 + c_3$.